\documentclass[aps,prxquantum,onecolumn,superscriptaddress,longbibliography]{revtex4-2}
\pdfoutput=1

\usepackage{amsmath,amssymb,amsfonts}
\usepackage{amsthm}
\usepackage{physics}
\usepackage{braket}
\usepackage{bbold}
\usepackage{graphicx}
\usepackage{float}
\usepackage{subcaption}
\usepackage{caption}

\usepackage{multirow}
\usepackage{booktabs}
\usepackage{algorithmicx}
\usepackage{algpseudocode}
\usepackage{xcolor}
\usepackage{tcolorbox}
\usepackage{xcolor}
\definecolor{purple}{rgb}{0.8,0,0.6}

\usepackage{siunitx}
\usepackage{upgreek}
\usepackage[english]{babel}
\usepackage{hyperref}
\usepackage[title]{appendix}

\begin{document}

\title{Cosmological Pseudo-Entropy}

\author{Manghang Limbu}
\affiliation{St. Xavier’s College, Tribhuvan University, Maitighar, 44600 Kathmandu, Nepal}
\affiliation{Holographic Himalaya, Lambagar, 44600 Tarakeshwar, Nepal}

\author{Pramod Kamal Kharel}
\affiliation{Holographic Himalaya, Lambagar, 44600 Tarakeshwar, Nepal}
\affiliation{Department of Physics and Astronomy, Ohio University, Athens, OH 45701, USA}

\author{Rohit Bhattarai}
\affiliation{Holographic Himalaya, Lambagar, 44600 Tarakeshwar, Nepal}
\affiliation{Tri-Chandra Multiple Campus, Tribhuvan University, Ghantaghar, 44600 Kathmandu, Nepal}

\author{Kiran Adhikari}
\email{kiran.adhikari@tum.de}
\affiliation{Emmy Noether Group for Theoretical Quantum Systems Design, Technical University of Munich, Arcisstraße 21, 80333 München}

\begin{abstract}
We study pseudo entropy $\mathcal{S}$, a recent generalization of entanglement entropy, for scalar cosmological perturbations in de Sitter space with sound speed $0.024 \leq c_s \leq 1$, and in expanding and contracting FLRW backgrounds with varying equation-of-state parameter $w$. In de Sitter space, $\mathrm{Re}(\mathcal{S})$ grows after horizon exit while
$c_s$ controls its onset and saturates at late times. A similar saturation occurs in expanding-accelerating and contracting-decelerating backgrounds. In contrast, expanding-decelerating and contracting-accelerating backgrounds show large early-time $\mathrm{Re}(\mathcal{S})$ followed by oscillations after horizon re-entry. This happens because while the squeezing freezes, the squeezing angle doesn't. Unlike entanglement entropy, pseudo entropy possesses an imaginary part, $\mathrm{Im}(\mathcal{S})$, as well, which can encode the relative phase. $\mathrm{Im}(\mathcal{S})$ decays to zero in de Sitter and expanding-accelerating cases, but forms dense sub-Hubble oscillation bands in expanding-decelerating and contracting-accelerating backgrounds. Compared with entanglement entropy, Krylov complexity, and Nielsen circuit complexity, pseudo entropy captures otherwise hidden phase information; in the unsaturated regime, its slope is $\sqrt{2}$ times that of Nielsen complexity. Unlike circuit complexity, whose saturation bound is $w$-independent, pseudo entropy is sensitive to $w$ during the transition regime, making it a finer information theoretic diagnostic of cosmological dynamics.

\end{abstract}

\maketitle

\section{Introduction}
\label{sec:intro_pseudo}

Recently, there has been a surge in the applications of quantum information in high-energy physics, condensed matter physics, and cosmology. One of the most important tools used in this field is the entanglement entropy \cite{Horodecki:2009zz, Nielsen:2012yss, Vidal:2002rm, Bombelli:1986rw, Srednicki:1993im, Calabrese:2004eu, PhysRevLett.96.110404, Adhikari:2022whf, Adhikari:2025vtx, Adhikari:2026srf}. Entanglement entropy is defined as the von Neumann entropy of the reduced density matrix of a subsystem and quantifies the degree of quantum entanglement between the subsystem and its complement. In other words, entanglement entropy characterizes the amount of information encoded in nonlocal quantum correlations, reflecting the information of the total system that is inaccessible from local observations alone \cite{PhysRevA.53.2046, Adhikari:2021pvv, Adhikari:2025zoa}.

Recently, \cite{Nakata_2021} introduced a new quantity known as pseudo entropy in the context of holography to extend the geometric interpretation of entanglement entropy to time-dependent spacetimes, where a well-defined Euclidean minimal surface is absent. It is defined as the von Neumann entropy of a transition matrix constructed from two different quantum states: 
\begin{equation}
S\!\left(\mathcal{T}^{\psi|\varphi}_A\right) = -\mathrm{Tr}\left(\mathcal{T}^{\psi|\varphi}_A \log \mathcal{T}^{\psi|\varphi}_A\right), \qquad \mathcal{T}^{\psi|\varphi} = \frac{|\psi\rangle\langle\varphi|}{\langle\varphi|\psi\rangle}.
\end{equation}
 In subsequent works, holographic pseudo entropy has been interpreted as time-like entanglement entropy with its imaginary parts signalling the emergence of time in both AdS/CFT and dS/CFT \cite{Doi:2022iyj}. Recent research on pseudo entropy has rapidly expanded its definition beyond its holographic origins into diverse fields, including quantum field theory and many-body systems \cite{Mukherjee:2022jac,He:2023eap,Mollabashi:2020yie,Mollabashi:2021xsd,Adhikari:2025vdl}. Consequently, pseudo entropy has emerged as a fundamentally significant quantity in general quantum systems. For instance, as discussed in \cite{Nakata_2021}, it can be interpreted as quantifying the number of Bell pairs that can be extracted through a post-selection procedure for a certain class of transition matrices. This work has linked pseudo entropy to operational tasks in quantum information theory, including entanglement distillation processes. Similarly, pseudo entropy has also been extended to topological phases through the notion of topological pseudo entropy, introduced as a generalization of topological entanglement entropy in \cite{Nishioka:2021cxe, Caputa:2024qkk}. It has also been found that pseudo entropy can serve as a new quantum order parameter that distinguishes whether two states are in the same quantum phase \cite{Mollabashi:2020yie}. Although there has been a significant number of studies of entanglement entropy for cosmological perturbations, we found no work on pseudo entropy in this context.

Inflationary cosmology predicts the sets of perturbations on the Friedmann-Lemaitre-Robertson-Walker (FLRW) background metric \cite{Guth:1980zm}. These perturbations result from the amplification of quantum fluctuations during inflation into macroscopic perturbations that later evolve into large-scale structures such as galaxies, clusters, and cosmic microwave background anisotropies~\cite{Mukhanov:1982nu}. These cosmological perturbations can be described in the language of a two-mode squeezed state \cite{MUKHANOV1992203, Albrecht:1992kf}. The squeezed state formalism allows the straightforward study of cosmological perturbations in quantum information theory ~\cite{Brandenberger:1990bx, Grishchuk:1990bj, Bhattacharyya:2020kgu, Adhikari:2022oxr, Kharel:2025lek}. 

Our goal in this paper is to study pseudo entropy in de Sitter Cosmology with different effective sound speeds to gain quantum information-theoretic insights into cosmological evolution and structure formation. We further generalize our analysis to Friedmann-Lemaitre-Robertson-Walker (FLRW) backgrounds with varying equation-of-state parameter $w$, in both expanding and contracting regimes. We found that pseudo entropy depends on both the squeezing amplitude and the squeezing phase. This explains the qualitative features we encountered across de Sitter and the expanding and contracting FLRW (accelerating and decelerating) regimes. We also found an exact relation between the growth rate of the real part of pseudo entropy and the equation of state in the transition regime across all four background classes. The effective sound speed only shifts the onset of events.

In section \ref{sec:pseudo}, we review some important details of pseudo entropy, especially its definition, properties, and interpretation. In section \ref{sec:Cosmo}, we review the description of cosmological perturbations as two-mode squeezed states. Thereafter, we compute the pseudo entropy between two squeezed-vacuum states and compare it with entanglement entropy, Nielsen's geometric complexity, and Krylov complexity. Further in section \ref{sec:de_sitter_app}, we discuss the pseudo entropy of cosmological perturbations in de Sitter cosmology with effective sound speed, and we also analyze the pseudo entropy of scalar perturbations in different backgrounds with various equations of state $w$. Finally, we mention the potential future directions in section \ref{sec:Outlook_pseudo} and conclude our work with all the cumulative major findings in section \ref{sec:Conc_pseudo}.


\section{Pseudo Entropy}
 \label{sec:pseudo}
Pseudo entropy can be thought of as a generalization of entanglement entropy to a transition between the two pure quantum states $|\psi\rangle$ and $|\varphi\rangle$ satisfying $\langle \varphi | \psi \rangle \neq 0$ \cite{ Nakata_2021, Mollabashi:2021xsd}. First, we define the transition matrix, $\mathcal{T}^{\psi|\varphi}$, as follows:
\begin{equation}
  \mathcal{T}^{\psi|\varphi} = \frac{|\psi\rangle\langle\varphi|}{\langle\varphi|\psi\rangle},
  \label{eq:transition}
\end{equation}
which is normalized such that its trace is unity, $   \mathrm{Tr} \left( \frac{|\psi\rangle\langle\varphi|}{\langle\varphi|\psi\rangle} \right) = \frac{\langle\varphi|\psi\rangle}{\langle\varphi|\psi\rangle} = 1$. Similarly, we find for any $n \in \mathbb{N}^+$,

\begin{align}
(\mathcal{T}^{\psi|\varphi})^n
&= \left( \frac{|\psi\rangle\langle\varphi|}{\langle\varphi|\psi\rangle} \right)^n  = \frac{|\psi\rangle\langle\varphi|}{\langle\varphi|\psi\rangle} = \mathcal{T}^{\psi|\varphi}.
\end{align}
Thus, $\mathrm{Tr}\left[\left(\mathcal{T}^{\psi|\varphi}\right)^n\right] = 1.$ Similarly, under the exchange of $|\psi\rangle$ and $|\varphi\rangle$, $  \mathcal{T}^{\varphi|\psi} = \left(\mathcal{T}^{\psi|\varphi}\right)^\dagger.$

In the literature, in order to compute the entanglement entropy, the total Hilbert space $\mathcal{H}_{\text{tot}}$ is divided into two parts $A$ and $B$, $  \mathcal{H}_{\text{tot}} = \mathcal{H}_A \otimes \mathcal{H}_B.$ Accordingly, the reduced transition matrix of system $A$ is defined by tracing out the sub-system $B$, 
\begin{equation}
  \mathcal{T}^{\psi|\varphi}_A \equiv \mathrm{Tr}_B\left[\mathcal{T}^{\psi|\varphi}\right] = \mathrm{Tr}_B\left[\frac{|\psi\rangle\langle\varphi|}{\langle\varphi|\psi\rangle}\right].
  \label{eq:reduced}
\end{equation}
and pseudo entropy is defined as:
\begin{equation}
\label{eq:Pseudo_entropy}
    S\!\left(\mathcal{T}^{\psi|\varphi}_A\right) = -\mathrm{Tr}\left(\mathcal{T}^{\psi|\varphi}_A \log \mathcal{T}^{\psi|\varphi}_A\right).
\end{equation}
In the case when the initial and final states are the same, i.e. $\ket{\psi} = \ket{\varphi}$, the pseudo entropy reduces to the ordinary entanglement entropy. Although equation \ref{eq:Pseudo_entropy} looks like Von-Neumann entropy, it can be complex as $\mathcal{T}^{\psi|\varphi}_A$ is, in general, non-Hermitian with complex eigenvalues. Pseudo entropy is real only in the special choice of the initial and final states. In contrast, Von Neumann entropy is always real. Operationally, pseudo entropy can be thought of as a measure of quantum entanglement for the intermediate states between the initial and final states \cite{Nakata_2021}.

For calculation purposes, it is convenient to define the $n$-th R\'enyi entropy of the transition matrix $\mathcal{T}^{\psi|\varphi}_A$ just like we define the $n$-th R\'enyi entropy of a quantum state $\rho$:
\begin{equation}
  S^{(n)}\!\left(\mathcal{T}^{\psi|\varphi}_A\right) \equiv \frac{1}{1-n}\log\mathrm{Tr}\!\left[\left(\mathcal{T}^{\psi|\varphi}_A\right)^n\right],
  \label{eq:renyi}
\end{equation}
where $(n \in \mathbb{N}^+,\, n \geq 2)$, and we can simply choose the branch of the log function: $-\pi < \mathrm{Im}[\log(z)] \leq \pi$. We call this quantity $S^{(n)}(\mathcal{T}^{\psi|\varphi}_A)$ the pseudo $n$-th R\'enyi entropy, and taking the $n \to 1$ limit, one obtains the pseudo entropy:
\begin{equation}
  S\!\left(\mathcal{T}^{\psi|\varphi}_A\right) \equiv \lim_{n\to 1} S^{(n)}\!\left(\mathcal{T}^{\psi|\varphi}_A\right)
  = S\!\left(\mathcal{T}^{\psi|\varphi}_A\right).
  \label{eq:von}
\end{equation}  

Here are some basic properties of  Pseudo R\'enyi entropy. If $|\varphi\rangle$ has no entanglement, then $S^{(n)}(\mathcal{T}^{\psi|\varphi}_A)=0$.  If $\mathcal{T}^{\psi|\varphi}_A$ and all eigenvalues of $\mathcal{T}^{\psi|\varphi}_A$ are in $\mathbb{C}^-$, then $S^{(n)}(\mathcal{T}^{\psi|\varphi}_A)=S^{(n)}(\mathcal{T}^{\varphi|\psi}_A)^*$. Furthermore, $S^{(n)}(\mathcal{T}^{\psi|\varphi}_A)=S^{(n)}(\mathcal{T}^{\varphi|\psi}_B)$. It is also possible to define the following real-valued quantity, which can be particularly relevant for phase transition studies. 
\begin{align}
\Delta S^{(n)} \!\left(\mathcal{T}^{\psi|\varphi}_A\right) = \frac{1}{2} \Big[ &S^{(n)} \!\left(\mathcal{T}^{\psi|\varphi}_A\right) + S^{(n)} \!\left(\mathcal{T}^{\varphi|\psi}_A\right) -   S^{(n)}\!\left(\mathcal{T}^{\psi|\psi}_A\right) -   S^{(n)}\!\left(\mathcal{T}^{\varphi|\varphi}_A\right) \Big]
\end{align}
where, we note that $S^{(n)} \left(\mathcal{T}^{\psi|\varphi}_A\right) = S^{(n)} \left(\mathcal{T}^{\varphi|\psi}_A\right)^*$, and the latter two terms are the standard entanglement entropy for the state $\ket{\psi}$ and $\ket{\varphi}$ respectively. Taking the $n \to 1$ limit, we obtain $   \Delta S \left(\mathcal{T}^{\psi|\varphi}_A\right) =  \lim_{n\to 1} \Delta S^{(n)} \left(\mathcal{T}^{\psi|\varphi}_A\right).$
$ \Delta S \left(\mathcal{T}^{\psi|\varphi}_A\right)$ is thus nothing but the difference between the real part of the pseudo entropy and the averaged entanglement entropy:
\begin{align}
\Delta S \!\left(\mathcal{T}^{\psi|\varphi}_A\right) = &\,\text{Re} \!\left(S\!\left(\mathcal{T}^{\psi|\varphi}_A\right) \right) - \frac{1}{2} \left( S(\rho_\psi)_A +  S(\rho_\varphi)_A \right)
\end{align}
where $ S(\rho_\psi)_A$ and $S(\rho_\varphi)_A$ is the standard entanglement entropy for the state $\ket{\psi}$ and $\ket{\varphi}$ of subsystem $A$.

\section{ Pseudo entropy of Cosmological perturbations}
\label{sec:Cosmo}
In this section, we will first review the quantum cosmological perturbations. For a detailed review of cosmological quantum perturbations and quantum fields in curved spacetime, we refer to \cite{MUKHANOV1992203, Mukhanov:2007zz}. 

We consider scalar perturbations of the FLRW metric by introducing small deviations $\delta g_{\mu\nu}$ about the homogeneous and isotropic background. In the longitudinal gauge, the perturbed metric can be written as
\begin{align}
ds^2 = a^2(\tau)\big[ &-(1+2\psi(\mathbf{x},\tau))\,d\tau^2 + (1-2\psi(\mathbf{x},\tau))\,d\mathbf{x}^2 \big],
\end{align}
which describes scalar perturbations of a spatially flat FLRW universe in the absence of anisotropic stress. We employ the Mukhanov-Sasaki formalism to describe scalar perturbations of the inflationary background. In the longitudinal gauge, and for a single scalar field with vanishing anisotropic stress, the curvature perturbation may be written as $\mathcal{R} = \psi + \frac{H}{\dot{\varphi}_0}\,\delta\varphi,$
where the overdot denotes differentiation with respect to cosmic time $t$, and \(H=\dot a/a\). Expanding the Einstein-Hilbert action and the matter action to second order, and then using the background Friedmann equations together with the linearized constraint equations, the quadratic action can be written as \cite{MUKHANOV1992203}
\begin{equation}
\label{eq:cosmoAction}
S = \frac{1}{2}\!\int\! dt\,d^3x\,a^3\frac{\dot{\varphi}_0^2}{c_s^2 H^2}
\!\left[\dot{\mathcal{R}}^{\,2}-\frac{c_s^2}{a^2}(\partial_i\mathcal{R})^2\right]\!.
\end{equation}
Here, the effective sound speed is defined by $c_s=\sqrt{\frac{\dot p}{\dot \rho}},$
where \(p\) and \(\rho\) are the effective pressure and energy density, respectively. The canonical single-field slow-roll limit corresponds to \(c_s=1\), while models with \(c_s<1\) describe a broader class of non-canonical scalar-field theories. It is convenient to introduce the Mukhanov variable
$\nu \equiv z\,\mathcal{R},$
 where $z \equiv \frac{a\dot{\varphi}_0}{c_s H},$
or equivalently, in conformal time \(\tau\),
$\nu = z\,\mathcal{R},$ where $z = a\frac{\varphi_k^{'}}{\mathcal{H}c_s},$ 
where a prime denotes differentiation with respect to conformal time and \(\mathcal{H}=a'/a\). In terms of \(\nu\), the action in equation~\eqref{eq:cosmoAction} takes the standard quadratic form
\begin{equation}
S = \frac{1}{2}\int d\tau\,d^3x
\left[
\nu'^2 - c_s^2(\partial_i\nu)^2 + \frac{z''}{z}\nu^2
\right].
\end{equation}
The perturbation modes therefore evolve independently and satisfy a harmonic-oscillator equation with a time-dependent effective mass. In particular, the mode equation is
\begin{equation}
\nu'' - c_s^2\nabla^2\nu - \frac{z''}{z}\nu = 0,
\label{eq:h_oscillator}
\end{equation}
which in Fourier space reads $\nu_k'' + \left(c_s^2 k^2 - \frac{z''}{z}\right)\nu_k = 0$, and is mathematically a parametric oscillator. In inflationary backgrounds, however, \(z''/z\) typically grows monotonically rather than oscillating, so the system does not exhibit reheating-type parametric resonance. Instead, after horizon exit, the effective frequency becomes imaginary, leading to the monotonic amplification of one quadrature and hence two-mode squeezing.

\subsection{ Quantization and Squeezed states formalism}
 \label{squeeze}

Since the classical quadratic action is insufficient to capture the dynamics of linear perturbations, it is necessary to treat them quantum mechanically \cite{MUKHANOV1992203,Brandenberger:1993zc}. To make the quantization procedure transparent, it is convenient to work with an action whose second-order action can be written, up to a total derivative, as \cite{Albrecht:1992kf}
\begin{align}
\delta_2 S
= \frac{1}{2}\!\int\! d^4x \bigg[ &(v')^2 - c_s^2 (v_{,i})^2-2\frac{z'}{z}v'v +\left(\frac{z'}{z}\right)^2 v^2 \bigg].
\label{eq:working action}
\end{align}

After rewriting the Lagrangian in terms of \(\pi\), the corresponding Hamiltonian is
\begin{equation}
H
= \frac{1}{2}\int d^3x
\left[
\pi^2 + c_s^2 (v_{,i})^2 + 2\frac{z'}{z}v\pi
\right].
\label{eq:hamiltoninan 1}
\end{equation}

After promoting the field and its conjugate momentum to operators, expanding them in Fourier modes, and then again expanding the field operators in terms of ladder operators \(\hat c_{\mathbf{k}}\), the Hamiltonian in equation \ref{eq:hamiltoninan 1} yields, 

\begin{align}
\hat H
= \frac{1}{2}\!\int\! \frac{d^3k}{(2\pi)^3}
\bigg[ & c_s k\!\left( \hat c_{\mathbf{k}}\hat c_{\mathbf{k}}^\dagger + \hat c_{-\mathbf{k}}^\dagger\hat c_{-\mathbf{k}} \right)  -i\frac{z'}{z}\!\left( \hat c_{\mathbf{k}}\hat c_{-\mathbf{k}} - \hat c_{\mathbf{k}}^\dagger \hat c_{-\mathbf{k}}^\dagger \right) \bigg].
\label{eq:Hamiltonian}
\end{align}
The first term is the free harmonic-oscillator contribution with dispersion relation $\omega_k = c_s k$, whereas the second term encodes the interaction with the expanding background and is responsible for the creation of correlated \(\mathbf{k}\) and \(-\mathbf{k}\) pairs. The time evolution of the ladder operators can be expressed in terms of a Bogoliubov transformation, leading to
\begin{align}
\hat c_{\mathbf{k}}(\tau)
= &\, e^{i\theta_k(\tau)}\cosh r_k(\tau)\,\hat c_{\mathbf{k}}(\tau_0) 
 + e^{-i\theta_k(\tau)+2i\phi_k(\tau)}\sinh r_k(\tau)\,\hat c^\dagger_{-\mathbf{k}}(\tau_0),
\end{align}
where \(r_k(\tau)\), \(\phi_k(\tau)\), and \(\theta_k(\tau)\) denote the squeezing parameter, squeezing angle, and rotation angle, respectively. This result is obtained by solving the coupled Heisenberg equations of motion for the creation and annihilation operators in an expanding background \cite{Albrecht:1992kf}. This final expression will be central in the subsequent analysis, as it explicitly shows how the initial vacuum state evolves into a two-mode squeezed state.

The evolution operator for each mode can be factorized into a rotation operator and a two-mode squeezing operator $\hat{\mathcal U}_{\mathbf{k}}
=
\hat{\mathcal S}_{\mathbf{k}}(r_{\mathbf{k}},\phi_{\mathbf{k}})
\hat{\mathcal R}_{\mathbf{k}}(\theta_{\mathbf{k}}),$
where
\begin{align}
\hat{\mathcal R}(\theta_{\mathbf{k}})
&= \exp\!\bigg[ -i\theta_{\mathbf{k}} \Big( \hat c_{\mathbf{k}}^\dagger \hat c_{\mathbf{k}} + \hat c_{-\mathbf{k}}^\dagger \hat c_{-\mathbf{k}} +1 \Big) \bigg] \\
\hat{\mathcal S}(r_{\mathbf{k}},\phi_{\mathbf{k}})
&= \exp\!\bigg[ \frac{r_{\mathbf{k}}}{2} \Big( e^{-2i\phi_{\mathbf{k}}}\hat c_{-\mathbf{k}}\hat c_{\mathbf{k}}- e^{2i\phi_{\mathbf{k}}}\hat c_{\mathbf{k}}^\dagger \hat c_{-\mathbf{k}}^\dagger \Big) \bigg].
\label{eq:squeezing_operator}
\end{align}
Since the rotation operator changes only the overall phase, it will not play an essential role in what follows. Acting on the vacuum, the squeezing operator produces a two-mode squeezed state,
\begin{align}
\ket{SQ(k,\tau)}
&= \hat{\mathcal S}_{k}(r_k,\phi_k)\ket{0_k,0_{-k}} \nonumber \\
&= \frac{1}{\cosh r_k} \sum_{n=0}^{\infty} e^{-2in\phi_k}\tanh^n r_k \ket{n_k,n_{-k}},
\label{eq:Squeezed state}
\end{align}
where $\ket{n_k,n_{-k}}
=
\frac{1}{n!}
\left(\hat c_k^\dagger \hat c_{-k}^\dagger\right)^n
\ket{0_k,0_{-k}}.$
The squeezing parameters obey a coupled set of evolution equations \cite{Albrecht:1992kf}. In conformal time, these can be written as
\begin{align}
\frac{dr_k}{d\tau}
&= -\frac{a'}{a}\cos(2\phi_k),
\label{eq:squeezing_evolution1}\\
\frac{d\phi_k}{d\tau}
&= c_sk + \frac{a'}{a}\sin(2\phi_k)\coth(2r_k),
\label{eq:squeezing_evolution2}\\
\frac{d\theta_k}{d\tau}
&= c_sk + \frac{a'}{a}\sin(2\phi_k)\tanh r_k.
\end{align}
These equations determine the growth of squeezing for a given background scale factor \(a(\tau)\). For a stationary background, \(a' = 0\), no squeezing is generated from an initial vacuum state. In general, analytical solutions are available only for special cosmological backgrounds, such as de Sitter expansion; otherwise, the squeezing dynamics must be solved numerically.

\subsection{Pseudo Entropy Between Two Squeezed Vacuum States}

We consider two two-mode squeezed vacuum states of modes $(k,-k)$, with different squeezing parameters,  $|\psi\rangle = |r_1,\phi_1\rangle$ and $|\varphi\rangle = |r_2,\phi_2\rangle$. The state $\ket{r, \phi}$ has a standard Fock expansion
\begin{equation}
|r,\phi\rangle = \frac{1}{\cosh r}
\sum_{n=0}^{\infty} \left( e^{-2 i \phi} \tanh r \right)^n
|n_k, n_{-k} \rangle .
\end{equation}
For transition matrix, $\mathcal{T}^{\psi|\varphi} = \frac{|\psi\rangle \langle \varphi|}
{\langle \varphi | \psi \rangle}$, one needs to compute $\langle \varphi | \psi \rangle$ and $|\psi\rangle \langle \varphi|$. Let $q \equiv \tanh r_1 \, \tanh r_2 \, e^{-2 i (\phi_1 - \phi_2)}$, and the  inner product forms a geometric series:
\begin{align}
\langle \varphi | \psi \rangle
&= \frac{1}{\cosh r_1 \cosh r_2}
\sum_{n=0}^{\infty} q^n = \frac{1}{\cosh r_1 \cosh r_2} \frac{1}{1-q}.
\end{align}
which holds for $|q| \leq 1$. Using the expansions of $|\psi\rangle$ and $\langle\varphi|$, one obtains 

\begin{equation*}
|\psi\rangle \langle \varphi|
= \sum_{n,m=0}^{\infty}
\alpha_n \beta_m^* \, |n,n\rangle \langle m,m|, \quad \alpha_n = \frac{1}{\cosh r_1}
(\tanh r_1)^n e^{-2in\phi_1}, \quad \beta_m = \frac{1}{\cosh r_2}
(\tanh r_2)^m e^{-2im\phi_2}\,.
\end{equation*}
Let system $A$ denote $+k$ modes, and system $B$ denote $-k$ modes. For computing pseudo entropy of system $A$, we trace out the $-k$ from the transition matrix, 
\begin{equation}
\mathcal{T}^{\psi|\varphi}_A
= \sum_{n=0}^{\infty} w_n |n\rangle \langle n|,
\end{equation}
where $w_n =\frac{\alpha_n \beta_n^*}{\langle \varphi | \psi \rangle} = (1-q) q^n $ are the eigenvalues of the reduced operator, which can be a complex quantity. The pseudo entropy, $S(\mathcal{T}^{\psi|\varphi}_A)$, is now 
\begin{align}
S(\mathcal{T}^{\psi|\varphi}_A)
&= - \mathrm{Tr}\!\left[ \mathcal{T}^{\psi|\varphi}_A \, \log \mathcal{T}^{\psi|\varphi}_A \right] \nonumber \\
&= - \sum_{n=0}^{\infty} w_n \log w_n \nonumber \\
&= -\sum_{n=0}^{\infty} (1-q) q^n \left[ \log(1-q) + n \log q \right].
\end{align}
Using the geometric sums $\sum_{n=0}^{\infty} q^n = \frac{1}{1-q}$, $\sum_{n=0}^{\infty} n q^n = \frac{q}{(1-q)^2}$, fixing a branch (e.g., the principal branch),  we obtain the closed-form expression:
\begin{align}
S(\mathcal{T}^{\psi|\varphi}_A) = -\log(1-q) - \frac{q}{1-q} \log q , \qquad
q = \tanh r_1 \tanh r_2 \, e^{-2 i (\phi_1 - \phi_2)}.
\label{eq:pseudo_entropy_squeezed}
\end{align}

 \begin{figure}[htbp]
    \centering
    \includegraphics[width=0.6\linewidth]{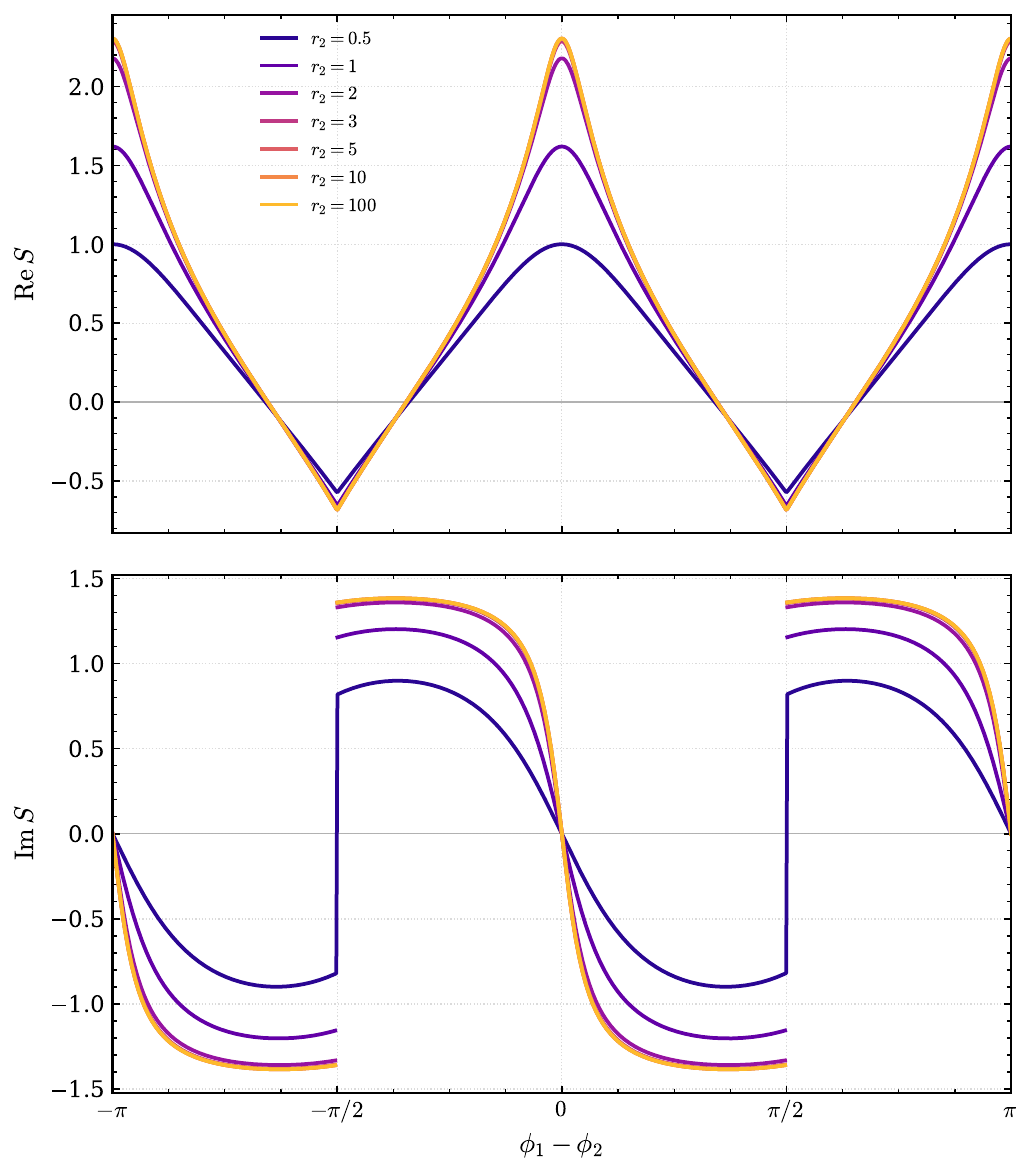}
    \caption{Real and imaginary parts of the pseudo entropy $S$ plotted as a function of $\phi_1 - \phi_2$ for fixed $r_1 = 1$ and various $r_2$. Curves saturate to a common envelope as $r_2 \gtrsim 3$.}
    \label{fig:pseudo-delta-phi}
\end{figure}

In figure~\ref{fig:pseudo-delta-phi}, we have plotted the real and imaginary parts of the pseudo entropy where several features of equation~\ref{eq:pseudo_entropy_squeezed} are visible. $\mathrm{Re}\,S$ is even, $\mathrm{Im}\,S$ is odd, and both are periodic in $\Delta\phi$ with period $\pi$. At $\Delta\phi = 0, \pm\pi$, $q$ is real and positive,
$T^{\psi|\varphi}_A$ is Hermitian, $\mathrm{Im}\,S$ vanishes, and $\mathrm{Re}\,S$ attains its maxima. At $\Delta\phi = \pm\pi/2$, $q$ lies on the negative real axis and $\log q$ crosses the principal branch cut, producing the cusps in $\mathrm{Re}\,S$ and the jumps in $\mathrm{Im}\,S$ seen in the figure; these are branch-dependent and not physical. The imaginary part is generically nonzero and reflects the non-Hermiticity of $\mathcal{T}^{\psi|\varphi}_A$, with no counterpart in ordinary entanglement entropy. Since the $r_2$-dependence enters only through $\tanh r_2$, the curves saturate rapidly: for $r_2 \gtrsim 3$ they are essentially indistinguishable from the $r_2 \to \infty$ envelope, and as either squeezing parameter goes to zero, $S$ vanishes smoothly.

When the two states are identical, $r_1 = r_2 = r$ and $\phi_1 = \phi_2 = \phi$, so that $q = \tanh^2 r \equiv \lambda \in (0,1)$ and the eigenvalues of $\mathcal{T}^{\psi|\psi}_A$, $w_n = (1-\lambda)\lambda^n, \qquad n = 0,1,2,\dots$ are real, positive, and normalized: $\sum_n w_n = 1$, the pseudo entropy becomes
\begin{align}
S = -\log(1-\lambda) - \frac{\lambda}{1-\lambda} \log \lambda = (\bar n + 1)\log(\bar n + 1) - \bar n \log \bar n
\label{eq:pseudo_entropy_reduced}
\end{align}
where, $\bar n = \frac{\lambda}{1-\lambda} = \sinh^2 r$ and $1+\bar n = \frac{1}{1-\lambda} = \cosh^2 r$, represents mean occupation number. $S$ is exactly the standard von Neumann entanglement entropy of a single mode of a two-mode squeezed vacuum. Hence, the pseudo entropy correctly reduces to the usual entanglement entropy when the states are identical.

\subsection{Initial state parameters}
\label{sec:intial_state_parameters}
In the next sections, we will compute the pseudo entropy of cosmological perturbations for various cosmological backgrounds. To compute pseudo entropy, we need the reference state. We choose the initial state as $|r_k^{(i)},\phi_k^{(i)}\rangle$ with 
$r_k^{(i)} = 10$ and $\phi_k^{(i)} = \pi$, and take the final state 
$|r_k(a),\phi_k(a)\rangle$ to be the analytically evolved mode at wavenumber $k = 0.01$, obtained by varying the scale factor $a$.

The pseudo entropy then tracks the transition between the two states as the universe evolves. These values are chosen for clarity rather than physical necessity. A large initial squeezing $r_k^{(i)} = 10$ saturates the reference state's contribution to the overlap, so that the $a$-dependence of $S$ reflects the evolved squeezing alone. In each background we fix the reference phase $\phi_k^{(i)}$ to coincide with
the late-time value of the evolved phase $\phi_k(a)$, so that the overlap $q = \tanh r_k^{(i)} \tanh r_k(a)\, e^{-2i(\phi_k^{(i)}-\phi_k(a))}$
approaches the positive real axis and the pseudo entropy attains a clean asymptotic limit ($\mathrm{Im}\,S \to 0$, $\mathrm{Re}\,S$ saturating). This is the reason for choice of phase $\phi_k^{(i)} = \pi$ in de sitter and $\phi_k^{(i)} = \pi/2$ in FLRW background. The wavenumber $k = 0.01$ places the horizon-crossing scale 
$a_\star = 2 c_s k \sim 10^{-2}$ near the centre of the plotted window 
$a \in [10^{-10}, 10^{10}]$, so that both the sub-horizon and super-horizon regimes are clearly visible.

\section{de Sitter Universe with various sound speeds}
\label{sec:de_sitter_app}

In this section, we will analyze the evolution of the pseudo entropy \( S \) in the de Sitter universe for various effective sound speeds. We chose this as our first application, because using the methods outlined in   \cite{Albrecht:1992kf}, the mode equations for the de Sitter case can be solved exactly:
\begin{align}
r_k(\tau) &= -\sinh^{-1}\!\left(\frac{1}{2 c_s k \tau}\right),
\label{rk_desitter} \\
\phi_k(\tau) &= -\frac{\pi}{4} - \frac{1}{2}\tan^{-1}\!\left(\frac{1}{2 c_s k \tau}\right).
\label{phik_desitter}
\end{align}
where $k$ is the co-moving wavenumber, $c_s$ denotes the sound speed of the perturbations, and $\tau$ represents the conformal time. Therefore, pseudo entropy can also be obtained analytically.   Physically, \( c_s = 1 \) corresponds to single scalar field slow-roll models, whereas \( c_s < 1 \) characterizes a broad class of non-canonical scalar field theories \cite{Choudhury:2021brg}. Cosmological observations constrain the effective sound speed to the allowed range \( 0.024 \leq c_s \leq 1 \) \cite{Planck:2015zfm}.

It is convenient to re-express equations~(\ref{rk_desitter}) and (\ref{phik_desitter}) in terms of the scale factor $a$ via $a(\tau) = -1/(H\tau)$. We now set $H = 1$, such that $a$ is measured in units of the Hubble radius. The horizon-crossing condition $|c_s k\tau| \sim 1$ then becomes the familiar $a \sim 2 c_s k$, and the limits $\tau \to -\infty$ 
and $\tau \to 0^-$ correspond to $a \to 0$ (deep sub-horizon) and 
$a \to \infty$ (deep super-horizon) respectively. The following observations are now clear. At early times, $a \ll 1$, and the modes lie well inside the horizon. In this regime, the squeezing parameter is negligible ($r_k(a) \ll 1$), and the squeezing angle remains approximately constant, $\phi_k(a) \approx -\pi/4$. As the universe expands, modes cross the horizon when $k \approx a$. Beyond horizon crossing, the system effectively behaves as an inverted harmonic oscillator and provides non-trivial contributions to the pseudo entropy. 

\subsubsection{Dependence on effective sound speeds}

\begin{figure}[t]
    \centering
    \captionsetup{font=small, labelfont=bf}
    \includegraphics[width=0.7\linewidth]{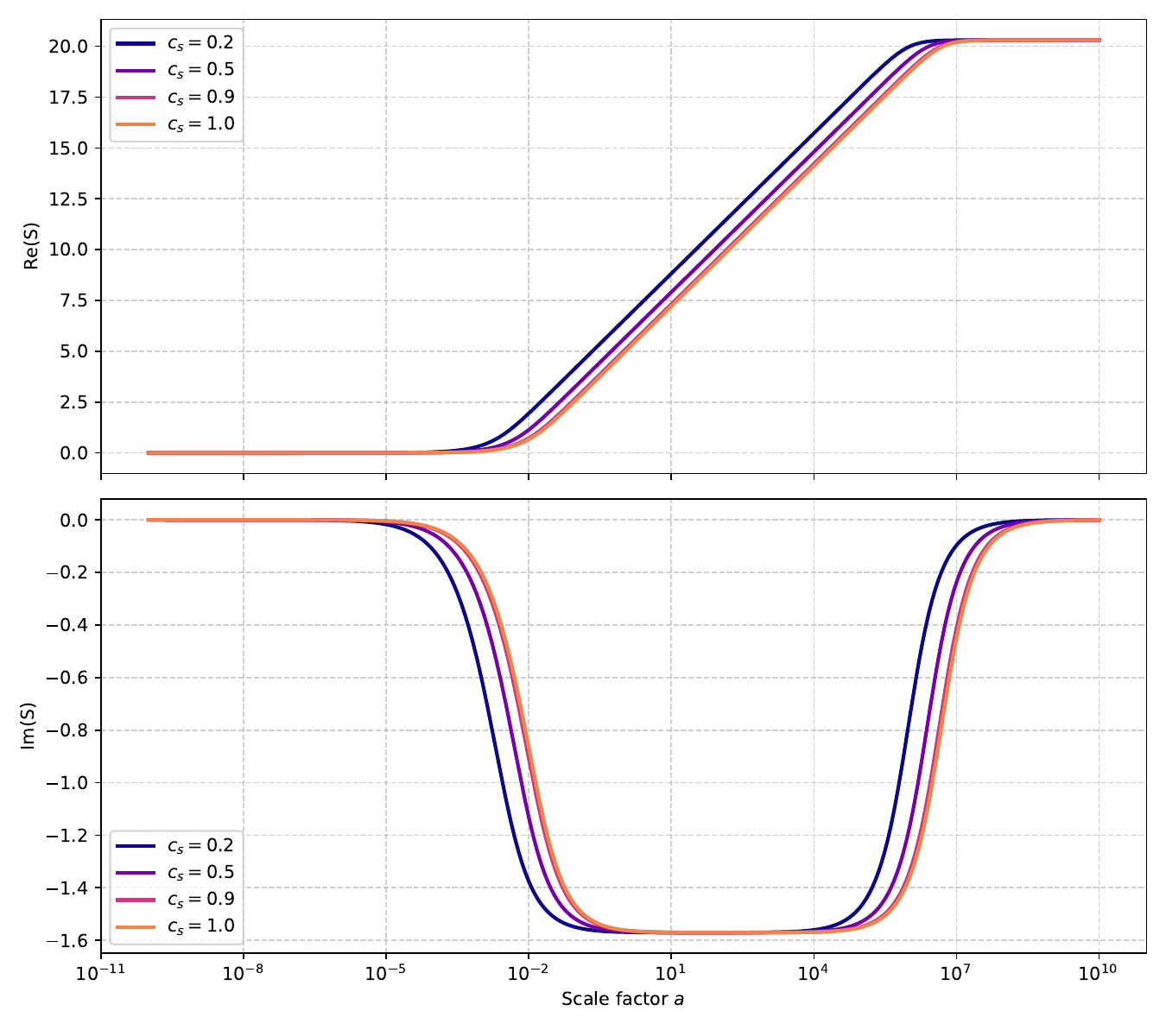}
    
    \caption{Real and imaginary parts of the pseudo entropy $S$ as a function of the scale factor $a$ in de Sitter background between a fixed reference state with parameters $r_k^{(i)} = 10$, $\phi_k^{(i)} = \pi$, and the evolved state. The mode $k = 0.01$ is fixed, while sound speeds are varied. }
    \label{fig:pseudo-deSitter-k-0.01}
\end{figure}

In figure \ref{fig:pseudo-deSitter-k-0.01}, we have plotted both real and imaginary parts of pseudo entropy for different values of sound speed $c_s$. In the late-time regime ($a \gg 1$), the real part of pseudo entropy grows logarithmically with the scale factor, corresponding to linear growth in cosmic time during de Sitter expansion, before eventually saturating at a value of approximately $\mathrm{Re}(S) \sim 20$. A useful interpretation of the real part of the pseudo entropy is that it quantifies the amount of quantum correlations shared between the paired modes ($k$ and $-k$) as the system evolves between two different times. The effect of varying the sound speed $c_s$ primarily shifts the onset of this growth. However, an important point to note is that, while the squeezing parameter and, consequently, the von Neumann entropy do not saturate at late times, the pseudo entropy does \cite{Kharel:2025lek}. From the equation of pseudo entropy for two-mode squeezed states \ref{eq:pseudo_entropy_squeezed}, we can see that at late times, $\tanh r_k(a) \to 1$ while $\Delta \phi(a) \to \Delta \phi_\infty$, a constant. Therefore,
$q(a) \to e^{-2i\Delta \phi_\infty},$
which lies on the unit circle and no longer evolves with time. Consequently, the pseudo entropy approaches a constant value, $S \to S_\infty,$ determined entirely by the frozen phase difference. Thus, the late-time saturation of pseudo entropy is a physical effect: although the squeezing parameter continues to grow, the distinguishability between the initial and evolved states becomes time-independent once the phase freezes. 

Unlike entanglement entropy, pseudo entropy has a non-vanishing imaginary part that encodes phase information. We find that, for the de-Sitter background, the imaginary part exhibits a transient negative dip before returning to zero at late times. The depth of the dip remains approximately unchanged, while its position shifts with $c_s$, with smaller values leading to an earlier onset. This behavior reflects the role of $c_s$ in determining the horizon-crossing time and, consequently, the phase evolution of the modes.

\subsubsection{Dependence on the choice of initial state}

\begin{figure}[t]
    \centering
    \captionsetup{font=small, labelfont=bf}
    \includegraphics[width=0.7\linewidth]{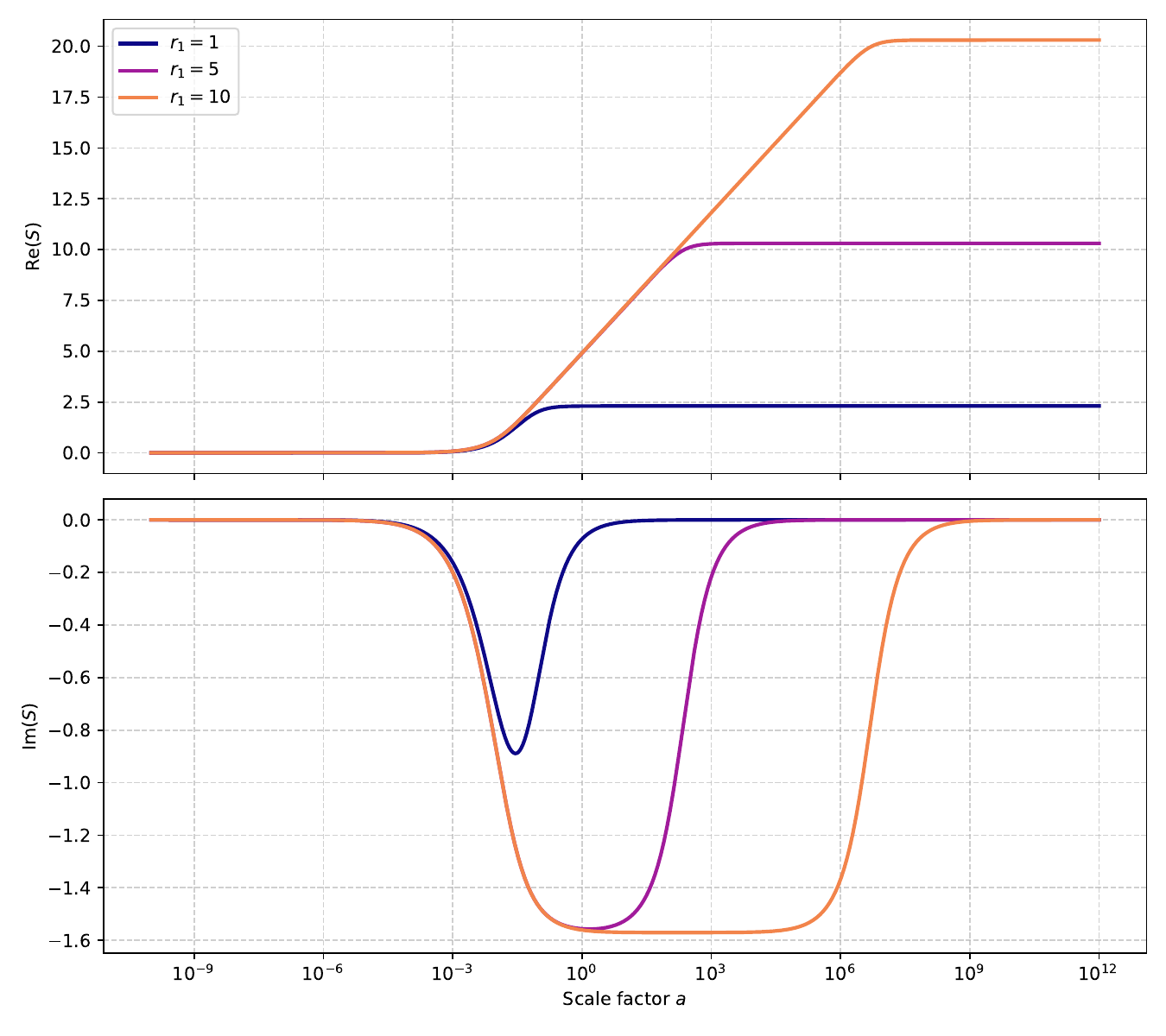}
    \caption{Pseudo entropy $S$ as a function of the scale factor $a$ in de Sitter background between a different initial squeezed state and the analytically evolved mode state, for fixed $k = 0.01$, and $c_s = 1.0$.}
    \label{fig:pseudo-deSitter_vary_r1}
\end{figure}

We also examine the dependence of pseudo entropy on the choice of the initial state by varying the initial squeezing parameter $r_k^{(i)}$ while keeping all other parameters fixed. Figure~\ref{fig:pseudo-deSitter_vary_r1} shows the evolution for $r_k^{(i)} = 1, 5, 10$ with $\phi_k^{(i)} = \pi$ for the mode $k = 0.01$ in de Sitter background. At early times ($a \ll 1$), all curves coincide, and the pseudo entropy remains negligible, indicating that initial-state dependence is suppressed deep inside the horizon. As the modes approach horizon crossing ($k \sim a$), the pseudo entropy begins to grow largely independent of the initial state. However, the magnitude of saturation shows a clear dependence on $r_k^{(i)}$, with $\mathrm{Re}(S)_{\mathrm{sat}} \propto r_k^{(i)}$, demonstrating that the pseudo entropy retains significant memory of the initial quantum state. This behavior arises because, for $r_k^{(i)} = 5, 10$, we have $\tanh r_k^{(i)} \approx 1$, whereas for $r_k^{(i)} = 1$, the reduced prefactor $\tanh(1) \approx 0.76$ leads to a visibly smaller amplitude in equation ~\ref{eq:pseudo_entropy_squeezed}. Consequently, $S$ saturates at a plateau height of approximately $2r_k^{(i)}$. Furthermore, whether $S$ saturates at early or late values of $a$ also depends on $r_k^{(i)}$: larger values of $r_k^{(i)}$ lead to freezing at later $a$, while smaller values result in freezing at earlier $a$. This is because the evolved squeezing grows logarithmically with the scale factor, $r_k(a)\sim \ln(a/c_s k)$ in equation \ref{rk_desitter}, so one unit of cosmological squeezing corresponds to one $e$-fold of super-horizon evolution. Saturation requires the evolved state to be squeezed at least as much as the reference state, $r_k(a) \geq r_k^{(i)}$, which translates into a delay of roughly $r_k^{(i)}$ $e$-folds past horizon crossing before $S$ saturates. Hence $r_k^{(i)} = 1$ saturates almost immediately after horizon exit, whereas $r_k^{(i)} = 10$ postpones saturation by roughly ten additional $e$-folds. The imaginary part exhibits a transient negative dip whose depth increases and whose position shifts to later times with increasing $r_k^{(i)}$, reflecting the enhanced phase mismatch for larger initial squeezing. 

\section{FLRW Universe: Expanding and Contracting backgrounds}
\label{ex_con_bg}

In this section, we will now analyze the evolution of the pseudo entropy \( S \) in FLRW universes with scalar curvature perturbations across different cosmological backgrounds. For convenience, we transform from conformal time to physical time using the relation \( d\tau = \frac{dt}{a(t)} \), as done previously for de Sitter space. The resulting differential equations governing the squeezing parameters \( r_k \) and \( \phi_k \) for $c_s = 1$ are given by~\cite{Bhattacharyya:2020kgu}.
\begin{align}
    \frac{d r_k}{d a} &= -\frac{1}{a} \cos(2\phi_k)\,, \label{eq:squeezing_parameter} \\
    \frac{d \phi_k}{d a} &= \frac{k}{a \mathcal{H}} + \frac{1}{a} \coth(2r_k) \sin(2\phi_k)\,
    \label{eq:squeezing_angle}
\end{align}
We numerically solve the squeezing parameters $r_k$ and $\phi_k$  using the fourth-order Runge-Kutta (RK4) method that will be used to calculate the pseudo entropy between two arbitrary two-mode squeezed vacua. The plots of the squeezing parameters $r_k$ and $\phi_k$ for different backgrounds have been thoroughly studied in \cite{Kharel:2025lek}. As in the de Sitter case, we fix the initial squeezing parameter to the same $r_k^{(i)} = 10$ and a different angle $\phi_k^{(i)} = \pi/2$, for computing the pseudo entropy. 
\subsection{Expanding background}
\label{sec:exp_bg}
The scale factor $a(\tau)$ for an expanding background is given by~\cite{Bhattacharyya:2020kgu}
\begin{equation}
a(\tau)=
\begin{cases}
\displaystyle\left(\frac{\tau_{0}}{\tau}\right)^{\beta},
& -\infty<\tau<0,\;\tau_{0}<0,\;\beta>0\;(w<-1/3)\quad\text{(accelerating)}\,,\\[8pt]
\displaystyle\left(\frac{\tau}{\tau_{0}}\right)^{|\beta|},
& 0<\tau<\infty,\;\tau_{0}>0,\;\beta<0\;(w>-1/3)\quad\text{(decelerating)}\,,
\end{cases}
\label{eq:scale-factor-exp}
\end{equation}
where $\beta = -\frac{2}{1+3w}.$ For the expanding background, the equations of motion corresponding to Eqs.~\ref{eq:squeezing_parameter} and \ref{eq:squeezing_angle} become
\begin{align}
\frac{d r_k}{d a} &= -\frac{1}{a}\cos(2\phi_k),
\label{rkdiffeq}\\
\frac{d \phi_k}{d a} &=
k\frac{|\tau_0|}{|\beta|}\frac{1}{a^{1+1/\beta}}
+\frac{1}{a}\coth(2r_k)\sin(2\phi_k),
\label{pkdiffeq}
\end{align}
where the first term in Eq.~\ref{pkdiffeq} is the driving term and the second term is the geometric term.
\begin{figure*}[t]
    \centering
    \captionsetup{font=small, labelfont=bf}

    \begin{subfigure}[t]{0.49\linewidth}
        \centering
        \includegraphics[width=\linewidth]{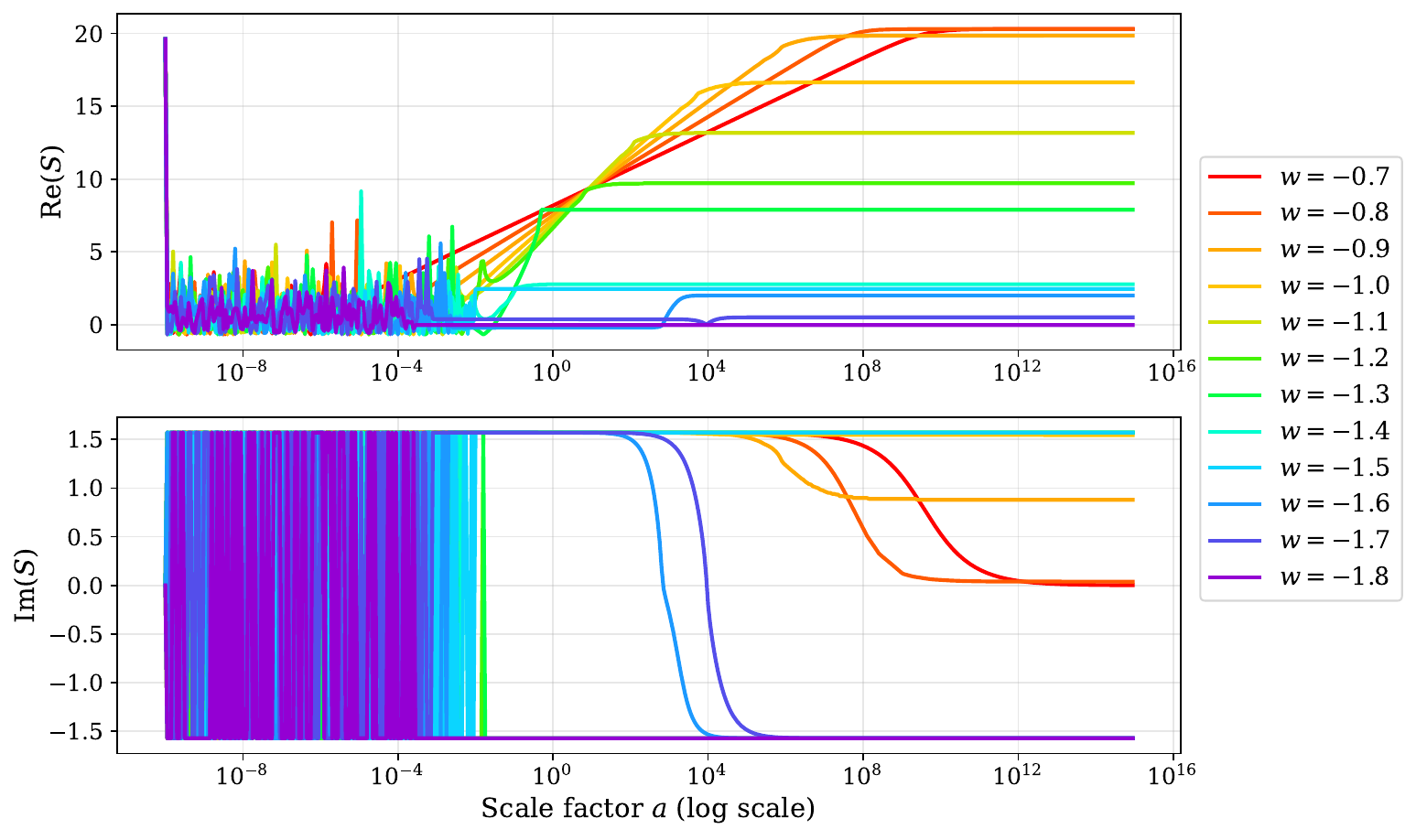}
        \caption{Accelerating case}
        \label{fig:expanding_accel_var_w2}
    \end{subfigure}
    \hfill
    \begin{subfigure}[t]{0.49\linewidth}
        \centering
        \includegraphics[width=\linewidth]{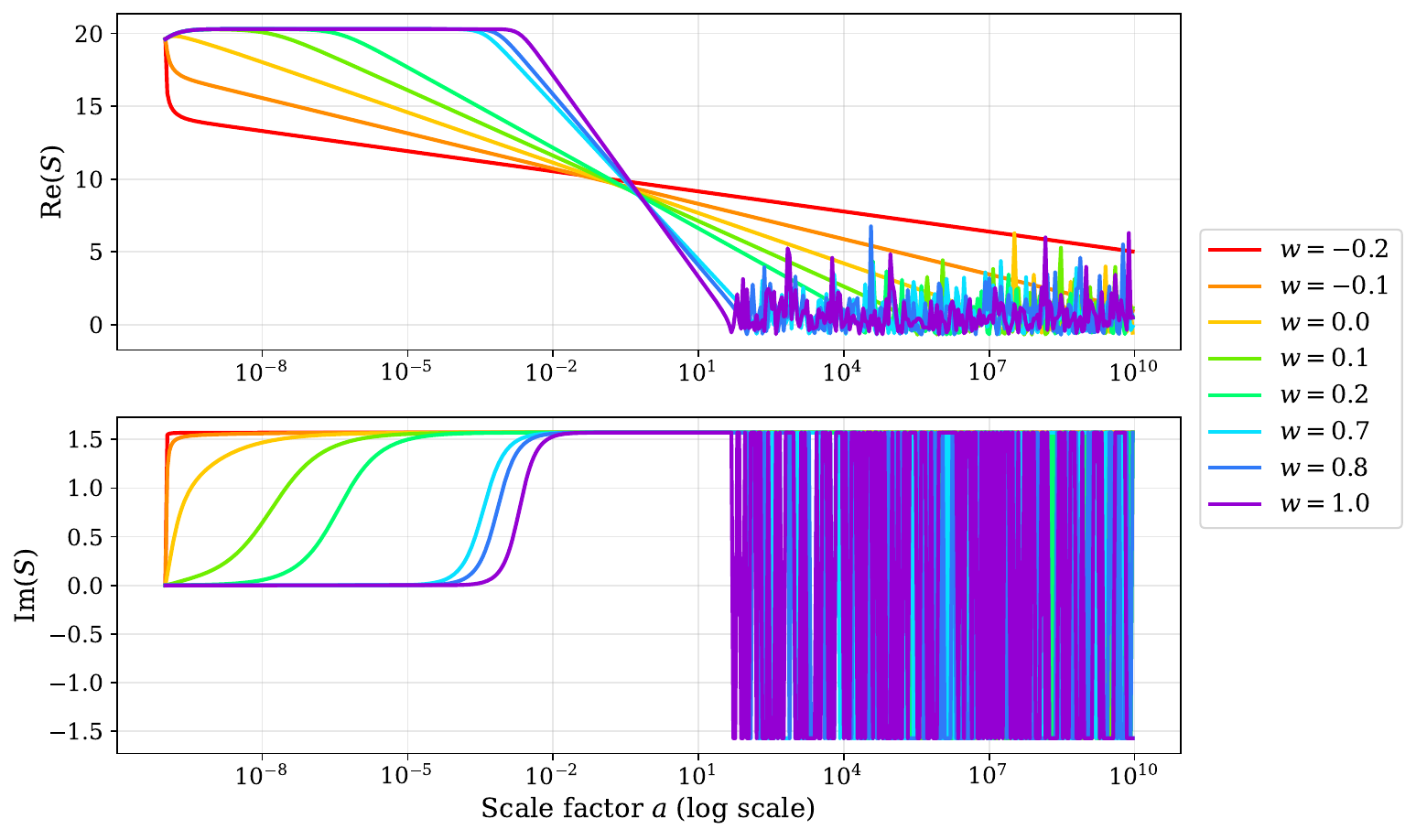}
        \caption{Decelerating case}
        \label{fig:expanding_dec_var_w2}
    \end{subfigure}

 \caption{pseudo entropy $S$ as a function of the scale factor $a$ for expanding backgrounds with fixed $k=0.001$, $r_k^{(i)}=10$, and $\varphi_k^{(i)}=-\pi/2$.}
    \label{fig:expanding_var_w}
\end{figure*}
\subsubsection{Accelerating case: }
First, we consider an expanding, accelerating background with a varying equation-of-state parameter $w<-1/3$, keeping $k=0.001$ fixed. Since $\beta>0$, the exponent of the scale factor in the driving term is
\begin{equation}
\alpha = 1+\frac{1}{\beta}>1,
\end{equation}
implying that the driving term decays faster than the geometric contribution, which scales as $a^{-1}$.

At early times ($a\ll1$), the modes are sub-horizon, and the pseudo entropy oscillates around a negligible mean value. Different values of $w$ cross the horizon at different transition scales,
\begin{equation}
a_{\mathrm{trans}}
\sim
\left(\frac{k|\tau|}{\beta}\right)^{\beta}.
\end{equation}
After horizon exit, the pseudo entropy grows approximately linearly before saturating at late times, as discussed in Sec.~\ref{sec:de_sitter_app}.

Figure~\ref{fig:expanding_accel_var_w2} shows that all curves converge near $a\sim10$. Beyond this point, the pseudo entropy saturates at $w$-dependent values. For example, for $w=-1.3$, the entropy reaches a plateau much earlier, while for $w=-0.7$ ($\alpha=1.55$), the slower decay of the driving term delays the saturation of $S$. This behavior originates from the earlier freezing of the phase after horizon exit for smaller values of $w$, such that the term $q \to e^{-2i\Delta\phi_\infty}$ becomes effectively time-independent earlier. Consequently, for $w<-1$ ($\alpha\ge2$), the rapid decay of the driving term suppresses further squeezing and leads to lower saturation entropy. The imaginary part correspondingly evolves from rapid oscillations to a frozen value of either $\pm\pi/2$ or $0$, determined by $\Delta\phi_\infty \pmod{\pi}$.

\subsubsection{Decelerating case: }

We now consider an expanding decelerating background with $w>-1/3$, keeping $k=0.001$ fixed. Since $\beta<0$, the driving exponent
\begin{equation}
\alpha = 1+\frac{1}{\beta}<1
\end{equation}
is smaller than unity. Hence, the driving term decays more slowly than the geometric term, which scales as $a^{-1}$, and grows with $a$ when $\alpha<0$.

At early times $a \ll 1$, modes are super-horizon and the field amplitude is frozen at $\phi_k(a) \approx \phi_k^{(i)}$, giving $q \approx 1$ and $\mathrm{Re}(S) \approx 2r_k^{(i)} \sim 20$. Each $w$ crosses the horizon at a different scale $a_{\mathrm{re\text{-}entry}}$. Figure~\ref{fig:expanding_dec_var_w2} shows that each curve transitions from constant to oscillatory behavior at these $w$-dependent re-entry times. The intersection near $a \sim 10^{-1}$--$10^{1}$ arises from staggered horizon re-entry, where curves at different stages of this transition momentarily overlap. Near the accelerating and decelerating boundary ($w = -0.2,\,-0.1$), $|\beta|$ is large, and $\alpha$ is only slightly less than unity, so the driving term decays almost as fast as the geometric term. The comoving Hubble radius therefore grows very slowly, modes take an extremely long time to re-enter, and even after re-entry begins, the driving term never strongly dominates over the geometric term. The squeezing built up during the super-horizon phase is eroded only gradually rather than being rapidly undone, which is why $\mathrm{Re}(S)$ decreases smoothly and slowly without sharp oscillations. For intermediate values ($w = 0.0$--$0.2$), modes re-enter at $a \sim 10^3$--$10^6$ and $\mathrm{Re}(S)$ decreases linearly with $\ln a$ before transitioning into oscillations. For stiff matter (\( w = 0.7\text{--}1.0 \)), \( \alpha < 0 \), and the driving term grows with the scale factor \( a \), causing modes to re-enter very early. As a result, the driving term quickly dominates over the geometric term, effectively undoing the squeezing: \( r_k \) freezes while \( \phi_k \) oscillates rapidly, driving \( q \) through all phases and pushing the pseudo entropy toward zero, where it subsequently oscillates. 
 
 The imaginary part rises smoothly from zero to $\pm\pi/2$ at $w$-dependent rates, followed by dense oscillation bands in the sub-horizon regime.

\subsection{Contracting Background}
\begin{figure*}[t]
    \centering
    \captionsetup{font=small, labelfont=bf}

    \begin{subfigure}[t]{0.49\textwidth}
        \centering
        \includegraphics[width=\linewidth]{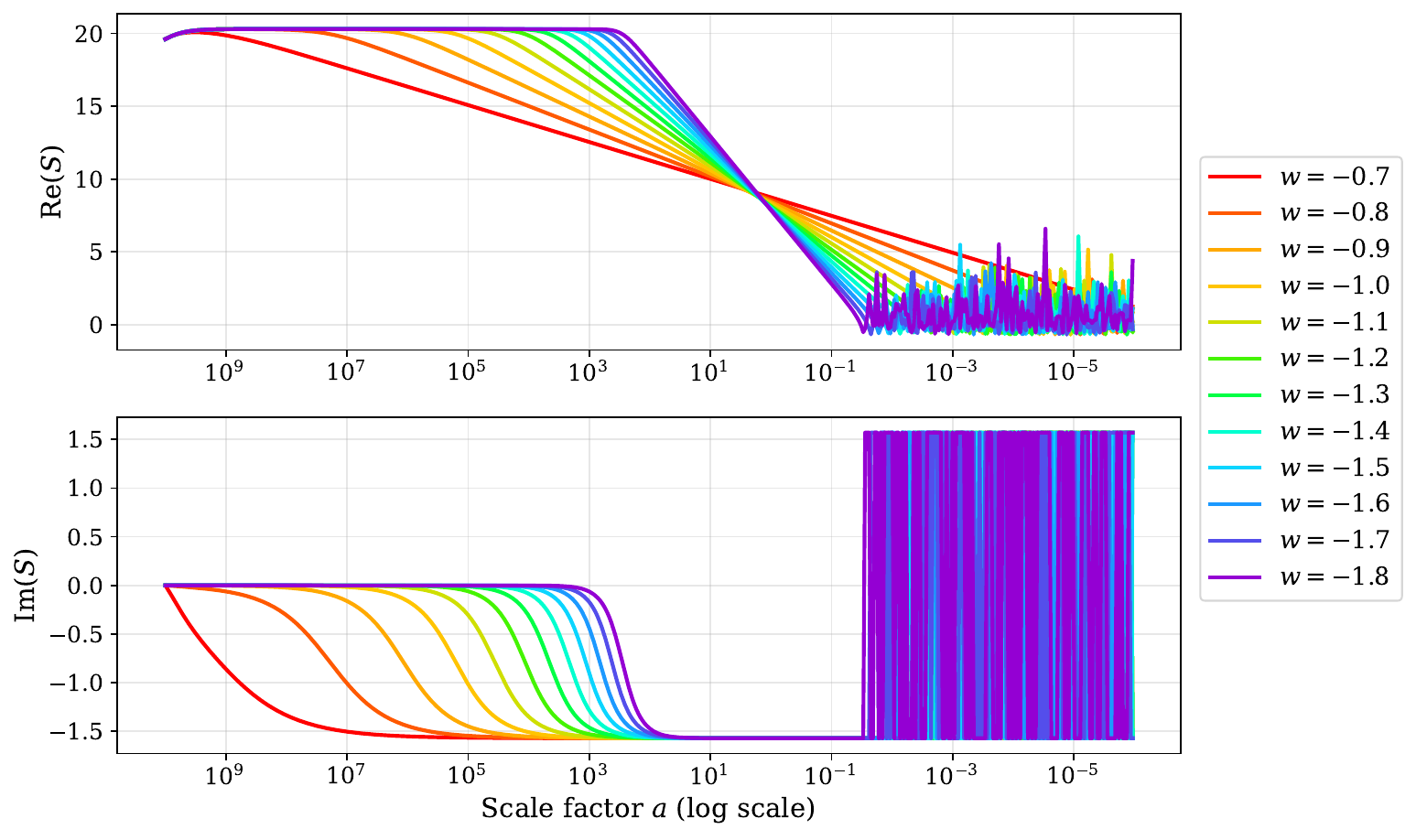}
        \caption{Accelerating case}
        \label{fig:pseudo_entropy_con_acc_vary_w}
    \end{subfigure}
    \hfill
    \begin{subfigure}[t]{0.49\textwidth}
        \centering
        \includegraphics[width=\linewidth]{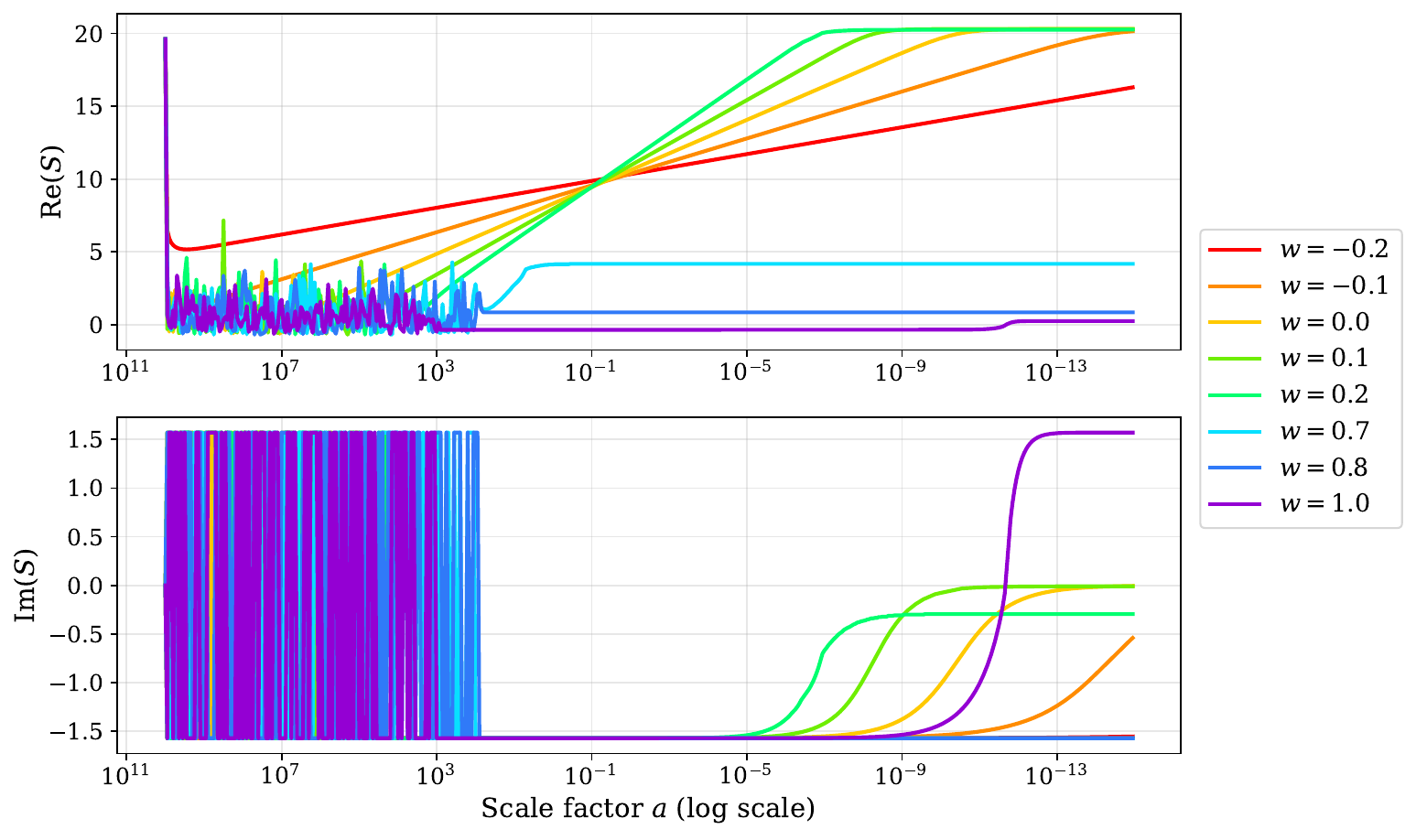}
        \caption{Decelerating case}
        \label{fig:pseudo_entropy_con_dec_vary_w}
    \end{subfigure}

    \caption{pseudo entropy $S$ as a function of the scale factor $a$ for contracting backgrounds with fixed $k=0.001$, $r_k^{(i)}=10$, and $\phi_k^{(i)}=\pi/2$. The $a$-axis is inverted to reflect the contracting evolution.}
    \label{fig:pseudo_entropy_con_var_w}
\end{figure*}

The  scale factor $a(\tau)$ for the contracting background is given by~\cite{Bhattacharyya:2020kgu}:
\begin{equation}
a(\tau)=
\begin{cases}
\displaystyle\left(\frac{\tau_{\rm 0}}{\tau}\right)^{\rm \beta},
& 0<\tau<\infty,\ \tau_{\rm 0}>0,\ \beta>0\;(w<-1/3)\quad\text{(accelerating)}\,,\\[8pt]
\displaystyle\left(\frac{\tau}{\tau_{\rm 0}}\right)^{\rm |\beta|},
& -\infty<\tau<0,\ \tau_{\rm 0}<0,\ \beta<0\;(w>-1/3)\quad\text{(decelerating)}\,,
\end{cases}
\label{eq:scale_factor_contr}
\end{equation}
where,  $\beta = -\frac{2}{1+3w}$. Then, the equations of motion equation~\ref{eq:squeezing_angle} are,
\begin{align}
\frac{d r_k}{d a} &= -\frac{1}{a} \cos(2\phi_k)\,,
\label{pkdiffeq12}
\\
\frac{d \phi_k}{d a} &=- k \frac{|\tau_0|}{|\beta|} \frac{1}{a^{\rm 1+1/\beta}} + \frac{1}{a} \coth(2r_k) \sin(2\phi_k)\,.
\label{pkdiffeq2}
\end{align}
We numerically solve the squeezing parameters $r_k$ and $\phi_k$  using the fourth-order Runge-Kutta (RK4) method. The plots of the solutions for both expanding and contracting backgrounds are provided in \cite{Kharel:2025lek}. We use numerically solved squeezing parameters from equation \ref{eq:pseudo_entropy_squeezed} to calculate pseudo entropy of cosmological perturbations. 

Given that a contracting universe is simply the time‐reverse of an expanding one, the pseudo entropy in a contracting accelerating background must mirror that in an expanding decelerating background. In particular, in an expanding, accelerating background, modes originate inside the horizon for early times $a \ll 1$ and gradually exit it as the universe evolves. In contrast, for a contracting universe, the behavior is reversed, with modes entering the horizon over time. Therefore, the squeezing solutions, and subsequently the pseudo entropy in a contracting accelerating background, are expected to resemble those found in expanding decelerating backgrounds. 
 
\subsubsection{Accelerating case:}
 
 For the contracting accelerating background with varying $w < -1/3$,  figure~\ref{fig:pseudo_entropy_con_acc_vary_w} shows that the curves descend to oscillatory, with $w = -1.3$ ($\alpha = 2.45$) dropping earliest and $w = -0.7$ ($\alpha = 1.55$) dropping very late. This is because the driving term scales as $a^{-\alpha}$, which grows as $a$ decreases; larger $\alpha$ (more negative $w$) causes the driving term to overwhelm the geometric term earlier during contraction, forcing the mode back into the sub-horizon regime sooner and undoing the squeezing at an earlier stage. For $w = -0.7$, $\alpha = 1.55$ is only slightly above unity, so the driving term grows slowly relative to the geometric term and the squeezing persists longer before being disrupted. All curves converge near $a \sim 0$--$10$ with a tighter spread than the expanding case; this is because all modes begin from the same frozen super-horizon state and descend together, with the fan-out developing only after the driving term begins to dominate at $w$-dependent scales, whereas in the expanding case, modes approach the frozen state from different initial sub-horizon conditions. At late times (small $a$), all modes have re-entered the horizon, $\phi_k$ oscillates rapidly, and pseudo entropy oscillates around small values. The imaginary part starts at zero and smoothly descends to $-\pi/2$ at staggered rates set by each $w$, followed by dense oscillation bands in the sub-horizon regime.

\subsubsection{Decelerating case:}

 Likewise, in the contracting decelerating case, the behavior closely mirrors that of an expanding accelerating background. At sufficiently early times ($a \ll 1$), the modes lie well inside the horizon and cross it around $k \approx a$, during which the squeezing parameter grows approximately linearly, similar to the expanding accelerating scenario. Consequently, the pseudo entropy of a contracting decelerating background exhibits a comparable evolution to that of an expanding accelerating background. For the contracting decelerating background with varying $w > -1/3$, figure~\ref{fig:pseudo_entropy_con_dec_vary_w} shows three distinct groups separated by the critical boundary $w = -1/3$ ($\alpha = 0$). The evolution of pseudo entropy is governed by the $w$-dependent driving term, which decays faster than the geometric term for $\beta > 0$ since $\alpha = 1 + \frac{1}{\beta} > 1$. At early times ($a \ll 1$), modes remain sub-horizon, and the pseudo entropy oscillates with negligible mean, while different $w$ values lead to horizon crossing at distinct scales. As the system evolves, all trajectories converge around $a \sim 10$, after which the pseudo entropy saturates to $w$-dependent values. In particular, smaller $w$ (especially $w < -1$) leads to faster decay of the driving term, earlier phase freezing, and consequently lower saturation entropy. The imaginary part shows dense oscillations at early times that settle to frozen values at $w$-dependent rates during the transition.

\section{Comparison to Other Measures}
In this section, we compare pseudo entropy with entanglement entropy,
Krylov complexity and Nielsen circuit complexity. The main distinction is
phase sensitivity: entanglement entropy and Krylov complexity depend only
on the squeezing amplitude $r_k$, Nielsen complexity depends on the
absolute squeezing phase $\phi_k$, while pseudo entropy depends on the
relative phase $\Delta\phi_k$ accumulated between two cosmological times. 

The entanglement entropy $S_{\mathrm{EE}}$ of the two-mode squeezed vacuum
is
\begin{align}
    S_{\mathrm{EE}} = (1+\sinh^2\!r_k)\ln(1+\sinh^2\!r_k)
           - \sinh^2\!r_k\,\ln\sinh^2\!r_k .
    \label{eq:EE}
\end{align}

The Krylov complexity $K$ for the two-mode squeezing operator was derived
in~\cite{Adhikari:2022oxr, Adhikari:2023evu} using the $SL(2,\mathbb{R})$ symmetry structure
of the Hamiltonian \cite{ Adhikari:2025vdl}. The operator wavefunctions in the Krylov basis are
$\varphi_n = e^{-2in\phi_k}\tanh^n\!r_k/\cosh r_k$, giving
\begin{equation}
    K = \sum_n n\,|\varphi_n|^2 = \sinh^2\!r_k .
    \label{eq:Krylov}
\end{equation}

Both entanglement entropy $S_{\mathrm{EE}}$ and Krylov complexity $K$
depend exclusively on the squeezing amplitude $r_k$. The squeezing angle
$\phi_k$, though present in the individual Krylov coefficients
$\varphi_n$, cancels exactly upon taking $|\varphi_n|^2$. This cancellation
reflects the fact that both measures are defined through traces or
modulus-squared operations that are invariant under phase rotations
$|\psi\rangle \to e^{i\theta}|\psi\rangle$~\cite{Baiguera:2025dkc}. As a
consequence, neither measure can detect the phase accumulated by the quantum
state during cosmological evolution.

The Nielsen geometric complexity of the cosmological squeezed state was
computed in~\cite{Bhattacharyya:2020rpy,Adhikari_2023}:
\begin{align}
C_{\mathrm{Nielsen}}
= &\underbrace{\left|\ln\left|\frac{1+z}{1-z}\right|\right|}_{\mathcal{T}_1}
  + \underbrace{\left|\tanh^{-1}\!\bigl(\sin(2\phi_k)\sinh(2r_k)\bigr)\right|}_{\mathcal{T}_2},
\qquad
z = \tanh r_k \, e^{-2i\phi_k}.
\label{eq:Nielsen}
\end{align}

Unlike $S_{\mathrm{EE}}$ and $K$, Nielsen complexity depends on the
absolute squeezing angle $\phi_k$ through the complex parameter
$z = \tanh r_k e^{-2i\phi_k}$. It therefore contains phase information that
is absent from the phase-blind measures.

However, Nielsen complexity also carries inherent ambiguities: its
numerical value depends on the choice of gate set, reference state, and
tolerance parameter~\cite{Adhikari:2022oxr, Baiguera:2025dkc}. Beyond these
conceptual issues, the two terms in Eq.~\eqref{eq:Nielsen} have different
mathematical properties in the cosmological setting. The first term
$\mathcal{T}_1$ is always well-defined since
$|(1+z)/(1-z)|$ is a positive real number for all $r_k$ and $\phi_k$.
The second term $\mathcal{T}_2$ contains $\tanh^{-1}$, which is real-valued
only when
\begin{equation}
    |\sin(2\phi_k)\sinh(2r_k)| < 1 .
\end{equation}
This bound is violated in the transition region $k|\tau|\sim 1$, where
$\phi_k \approx -\pi/4$ and $\sinh(2r_k)$ grows through unity, rendering
$\mathcal{T}_2$ ill-defined at the crossing point. In numerical implementations, this is handled by clipping the argument to
$[-1+\varepsilon,1-\varepsilon]$, which produces a visible dip in
$C_{\mathrm{Nielsen}}$ near the transition scale.

The pseudo-entropy of the two-mode squeezed vacuum is
\begin{align}
S(\mathcal{T}^{\psi|\varphi}_A)
= -\log(1-q) - \frac{q}{1-q}\log q,
\qquad
q = \tanh r_1 \tanh r_2 \,
e^{-2 i (\phi_1 - \phi_2)} .
\label{eq:pseudo_entropy_squeezed2}
\end{align}

There is some structural parallel between pseudo-entropy and Nielsen complexity. Both are built from the same squeezing data: $\tanh r_k$ as an amplitude weight and $e^{-2i\phi_k}$ as a phase factor. The key difference is that the parameter
\begin{equation}
    q = \tanh r_k^{(i)} \tanh r_k(a) e^{-2i\Delta\phi_k},
    \qquad
    \Delta\phi_k = \phi_k^{(i)} - \phi_k(a),
\end{equation}
appearing in pseudo entropy depends on the relative phase between two
cosmological times, whereas the Nielsen parameter
$z = \tanh r_k e^{-2i\phi_k}$ depends on the absolute squeezing phase
$\phi_k$. Pseudo entropy, therefore, directly measures phase information
accumulated during cosmological evolution from the reference time $\tau_0$
to the later time $\tau$.

This relative-phase dependence is the main distinction from both
phase-blind diagnostics and absolute-phase circuit complexity. While
$S_{\mathrm{EE}}$ and $K$ discard phase information, and Nielsen complexity
depends on a phase whose interpretation is tied to a choice of reference
state and circuit geometry, pseudo entropy compares two physical states, and
is sensitive to their relative phase separation.

\subsection{Cosmological Growth of Pseudo Entropy and Circuit Complexity}

The relative-phase sensitivity of pseudo entropy has a direct dynamical
consequence. In the transition region where $\mathrm{Re}(S)$ varies
linearly with $\ln a$, we define $ \lambda_S \equiv \frac{dS}{d(\ln a)} .$
From numerical fits across all values of $w$ and all four cosmological
backgrounds, we find
\begin{equation}
\lambda_S = \frac{1}{\beta} = -\frac{1+3w}{2},
\label{eq:lambda_S}
\end{equation}
with $R^2=1$ in every case. This result can be derived analytically. In the highly squeezed limit,
$\mathrm{Re}(S) \approx 1-\ln|2\Delta\phi|$, and the residual driving term
causes the phase difference to evolve as
$\Delta\phi \propto a^{-1/\beta}$. Therefore
\begin{equation}
    \mathrm{Re}(S) \approx \mathrm{const} + \frac{1}{\beta}\ln a .
\end{equation}
The slope is positive for accelerating backgrounds, $w<-1/3$, and negative
for decelerating backgrounds, $w>-1/3$. It is identical for expanding and
contracting backgrounds with the same value of $w$.

The circuit complexity slope for two-mode squeezed cosmological
perturbations was computed in~\cite{Bhattacharyya:2020kgu}. Converting to
the same variable $\ln a$, the complexity slope in the unsaturated regime is
\begin{equation}
    \frac{d\mathcal{C}}{d(\ln a)}
    =
    -\frac{1+3w}{2\sqrt{2}}
\end{equation}
for all four backgrounds. Hence, in the unsaturated regime,
\begin{equation}
\frac{|\lambda_S|}
{|d\mathcal{C}/d(\ln a)|}
=
\sqrt{2}.
\label{eq:ratio}
\end{equation}

This ratio is universal for all four backgrounds and all values of $w$
within the unsaturated range:
$-5/3 < w < -1/3$ for expanding accelerating backgrounds,
$-1/3 < w < 1$ for contracting decelerating backgrounds, and all $w$ for
the remaining two cases.

The distinction appears beyond the saturation threshold. Complexity
slope saturates at $w=-5/3$ for expanding accelerating backgrounds and at $w=1$
for contracting decelerating backgrounds, after which
$d\mathcal{C}/d(\ln a)=\pm \sqrt{2}$ becomes independent of $w$
\cite{Bhattacharyya:2020kgu}. This saturation gives rise to the bound
$d\mathcal{C}/dt \leq \sqrt{2}|H|$. Pseudo entropy's $ \lambda_S$, by contrast, continues
to scale as $|1+3w|/2$ in transition regime until late-time plateau because it tracks
$-\ln|\Delta\phi|$, with
$\Delta\phi = \phi_k^{(i)} - \phi_k(a)$. Since the reference state has
$\phi_k^{(i)}=\phi_\infty$, pseudo entropy continues to probe the residual
driving term $\propto a^{-1/\beta}$ and does not saturate. It can therefore
serve as a finer diagnostic of the equation of state in regimes where the
complexity slope has already saturated.

\subsection{Complementarity with entanglement entropy across cosmological backgrounds}

The relation between pseudo entropy and entanglement entropy depends on the
cosmological background. In the coincident limit, pseudo entropy reduces to
the standard entanglement entropy, as shown in
Eq.~\eqref{eq:pseudo_entropy_reduced}. Away from this limit, however,
$S_{\mathrm{EE}}$ depends only on the squeezing amplitude $r_k$, whereas
pseudo entropy is sensitive to the relative squeezing phase $\Delta\phi_k$.

In expanding-accelerating and de Sitter backgrounds, pseudo entropy
saturates at $\mathrm{Re}(S)\to S_\infty$, while $S_{\mathrm{EE}}$ continues
to grow without bounds, as discussed in
Sections~\ref{sec:de_sitter_app} and~\ref{ex_con_bg}. In decelerating
backgrounds the roles are reversed. For expanding-decelerating and
contracting-accelerating backgrounds, modes re-enter the horizon, and
$r_k$ freezes at late times~\cite{Kharel:2025lek}, so $S_{\mathrm{EE}}$
plateaus. Pseudo entropy, however, starts from
$\mathrm{Re}(S)\approx 2r_0$ due to the initial phase coherence
$\Delta\phi\approx 0$, decreases linearly with $\ln a$ with the universal slope $(\lambda_S)$ and ultimately oscillates in the sub-Hubble regime.
\begin{equation}
    \lambda_S = -\frac{1+3w}{2},
\end{equation}

The key point is that this slope depends explicitly on $w$ and takes the
same form in all four backgrounds, whereas the von Neumann growth rate
$dS_{\mathrm{EE}}/d(\ln a)\sim 2\,dr_k/d(\ln a)$ is approximately
$w$-independent across all four backgrounds~\cite{Kharel:2025lek}. This
universality follows directly from the squeezing equation 
\eqref{pkdiffeq12}, which has no explicit $w$-dependence. Once
$\phi_k$ has settled near the super-Hubble fixed point where
$\cos(2\phi_k)\to \pm 1$, one obtains
$dr_k/d(\ln a)\to \pm 1$ regardless of $w$; see Eq.~(83) of
Ref.~\cite{Kharel:2025lek}. The equation of state enters $r_k(a)$ only
through the horizon-crossing time and the duration of super-Hubble growth,
not through the slope itself.

Pseudo entropy therefore, provides a quantitative readout of the equation of
state that $S_{\mathrm{EE}}$ cannot provide. The two measures are best
understood as complementary: in backgrounds where one has saturated, the
other typically continues to evolve.

\section{Outlook}
\label{sec:Outlook_pseudo}

In this work, we have focused on cosmological perturbations within the linear regime, where different modes $\pm k$ evolve independently. Accordingly, we studied the pseudo entropy between a given mode \(k\) at an initial time and the same mode at a later time. While this setup allows for a controlled analysis, the physical interpretation of the imaginary part of pseudo entropy remains unclear and requires further investigation.\\
Closely related is a definition-level question about pseudo entropy that has not, to our knowledge, been addressed in previous work. The reduced transition matrix $\tau(a)$ in our setup has a geometric spectrum $\{(1-q)\,q^n\}$ controlled by the complex parameter $q(a)$
which appears explicitly in the simplified pseudo entropy formula~\ref{eq:pseudo_entropy_squeezed}
Along a cosmological trajectory, the Bogoliubov phase accumulates without bound and $q(a)$ winds around the origin of the complex plane, so that the complex logarithm $\log q$ requires an explicit branch choice that is logically distinct from the numerical evaluation of $S$. Previous studies have avoided this question in various ways: by restricting to Euclidean path-integral setups with real action~\cite{Mollabashi:2020yie}, or by working in settings where the transition matrix is pseudo-Hermitian ~\cite{Guo:2022jzs} which emerges from the causal structure of spacetime, as in Rindler-wedge operator insertions~\cite{Guo:2023tjv}. None of these options is available in our Lorentzian cosmological setup, where $q(a)$ is complex with non-trivial winding and the geometric spectrum $\{(1-q)q^n\}$ is generically not closed under complex conjugation, so that no pseudo-Hermitian structure of $\tau(a)$ is preserved along the evolution. We have adopted the principal branch $\arg z \in (-\pi, \pi]$, consistent with the default conventions underlying numerical evaluation and replica-style analytic continuation in the existing literature, and have reported the resulting discontinuities of $\mathrm{Im}S(a)$ as features of this convention. A systematic study of branch prescriptions for pseudo entropy along continuous Lorentzian trajectories, including whether the principal branch is singled out by a physical principle or merely by convention, and what operational content the alternative continuous-branch quantity may carry, is an important open problem. 

A natural extension of this work is to go beyond linear theory. In realistic cosmological settings, gravitational nonlinearities induce couplings between modes, particularly between sub-Hubble and super-Hubble modes \cite{Brahma:2021mng,Brahma:2020zpk}. Studying pseudo entropy in such interacting scenarios may reveal qualitatively new features absent in the linear approximation. Another direction is to study interacting quantum fields. Many realistic inflationary models, especially those motivated by string theory, naturally include additional scalar fields \cite{Turok:1987pg,Damour:2002vu,Kachru:2003sx,Krause:2007jk}. Investigating pseudo entropy in such setups could clarify how field interactions influence cosmological evolution. More broadly, pseudo entropy is a relatively new concept in quantum information theory and has not yet been extensively applied across different areas of physics. It has also been proposed as a quantity that can distinguish between quantum states and may serve as an indicator of quantum phase transition \cite{Nishioka:2021cxe, Caputa:2024qkk, Mollabashi:2020yie}. In this context, it would be worthwhile to investigate its behavior in systems exhibiting topological phase transitions, such as the Su-Schrieffer-Heeger (SSH) model and the Kitaev chain. Furthermore, it would be interesting to examine whether pseudo entropy can capture signatures of quantum chaos in many-body systems, including models such as the Heisenberg XXZ and Sachdev-Ye-Kitaev (SYK) models. Finally, motivated by developments in quantum information and holography, one may explore whether an analog of computational complexity can be defined for transition matrices associated with pseudo entropy \cite{Nakata_2021}. Such a construction, if well-defined, could provide a new way to characterize dynamical processes in quantum systems.

\section{Conclusion}
\label{sec:Conc_pseudo}
In this paper, we explored pseudo entropy as a probe of cosmological perturbations. It admits a compact, fully analytic description within the two-mode squeezed-state formalism of cosmological perturbation theory. The entire content of the reduced transition matrix between two squeezed vacua is captured by a single complex variable built from the two squeezing amplitudes and their relative phase. The resulting expression passes expected consistency checks, particularly collapsing to the familiar single-mode entanglement entropy when the two states coincide. This makes pseudo entropy a computationally tractable quantity in cosmology, on the same footing as entanglement entropy and circuit complexity.

Across all backgrounds we studied, the behavior of the pseudo entropy is controlled by a single mechanism: how the squeezing amplitude and phase evolve together. Whenever the background dynamics drives the phase to a frozen late-time value, the defining complex parameter of the transition matrix settles onto the unit circle, and the real part of the pseudo entropy stops growing, even if the squeezing itself does not. This single observation organizes all of the qualitative features we encountered across de Sitter, expanding and contracting FLRW, and accelerating and decelerating regimes. As a concrete quantitative outcome, we also found an exact, universal relation between the growth rate (slope) of the real part of the pseudo entropy in the transition regime and the equation of state, which holds without any fitting across all four background classes.

Pseudo-entropy is complementary to the quantum information diagnostics previously applied to cosmological perturbations. Measures that depend only on the squeezing amplitude are, by construction, blind to the phase dynamics that we have shown to control the late-time behavior. The existing measure that does see the phase, Nielsen circuit complexity, does so through the absolute squeezing angle and suffers from both conceptual ambiguities and a technical obstruction in the transition region. Pseudo entropy
instead depends on the relative phase between initial and
evolved states, which is the physically relevant quantity for
distinguishability, it is regular throughout the evolution, and yields an imaginary part with no analog in the other measures. The clean $\sqrt{2}$ ratio between the pseudo\,-\, pseudo entropy slope and the circuit complexity slope in the unsaturated regime, together with the fact that the pseudo entropy slope continues to track $w$ after the circuit complexity slope has saturated \cite{Bhattacharyya:2020kgu}, suggests that pseudo entropy can resolve features of cosmological dynamics that are invisible to existing complexity measures.

\acknowledgments
This research is part of the Abdus Salam International Centre for Theoretical Physics (ICTP): Physics Without Frontiers (PWF) initiative, and we acknowledge support from the PWF program of the ICTP, Italy. K.A. is supported by the Munich Quantum Valley, which is supported by the Bavarian state government with funds from the Hightech Agenda Bayern Plus.

\bibliographystyle{JHEP}

\bibliography{ref}
\end{document}